\pdfoutput=1
\documentclass[a4paper]{article}

\usepackage[cmex10]{amsmath}
\usepackage{amssymb}
\usepackage{latexsym}
\usepackage{indentfirst}

\usepackage{alltt}

\usepackage{subfigure}
\usepackage{amsthm}
\usepackage{graphicx}

\DeclareMathAlphabet{\mathcal}{OMS}{cmsy}{m}{n}

{\theoremstyle{definition}

}

\newcommand{\revision}[1]{{{\pdfliteral{0 0 1 rg}#1\pdfliteral{0 0 0 rg}}}}

\newcommand{\type}[1]{{\sf\small #1}}

\newcommand{\ie}{i.e.}

\newcommand{\forallt}{\textbf{for all}}
\newcommand{\foranyt}{\textbf{for any}}

\newcommand{\mrcv}[2]{\copr{receive}\;<\;\(#1(#2),#2\)\;>}

\newcommand{\msnd}[2]{\copr{send}\;<\;\(#1,#2\)\;>}

\newcommand{\orr}{\textbf{OR}}
\newcommand{\doo}{\textbf{do}}
\newcommand{\goes}{\rightarrow}

\newcommand{\cdef}[1]{#1}
\newcommand{\ccom}[1]{#1}
\newcommand{\ccon}[1]{#1}
\newcommand{\ccod}[1]{#1}
\newcommand{\copr}[1]{#1}

\newcommand{\emptyline}{\vspace{1ex}}

\usepackage{authblk}

\title{\textbf{Exploring Design Tradeoffs Of A Distributed Algorithm For Cosmic Ray Event Detection}}

\author[1]{Suhail Yousaf\thanks{s.yousaf@vu.nl}}
\author[1]{Rena Bakhshi}
\author[1]{Maarten van Steen}
\author[1]{Spyros Voulgaris}
\author[2]{John L. Kelley}
{
\affil[1]{\normalsize{The Network Institute, and Dept.\ of Computer Science, VU University Amsterdam, De Boelelaan 1081, 1081 HV Amsterdam, The Netherlands}}
\affil[2]{Radboud University Nijmegen, IMAPP / Department of Astrophysics, Heyendaalseweg 135, 6500 GL Nijmegen, The Netherlands}
}

\date{}

\begin{document}
 \maketitle

\begin{abstract}
Many sensor networks, including large particle detector arrays
  measuring high-energy cosmic-ray air showers, traditionally rely on
  centralised trigger algorithms to find spatial and temporal coincidences
  of individual nodes.  Such schemes suffer from scalability problems, especially if
  the nodes communicate wirelessly or have bandwidth limitations.
  However, nodes which instead communicate with each other can, in
  principle, use a distributed algorithm to find coincident events
  themselves without communication with a central node.  We present such an
  algorithm and consider various design tradeoffs involved, in the context of a
  potential trigger for the Auger Engineering Radio Array (AERA).
\end{abstract}



\section{Introduction}
\label{Sec:Intro}

There is an increasing trend in monitoring large spatial areas, such as 
forests for fire detection, mountains and hills for landslide and avalanches, 
and volcanoes for early warning or scientific study. This class of detectors 
also includes experiments measuring cosmic rays, high energy particles 
from space whose origin and acceleration mechanisms are still 
under debate.  
For example, the Pierre Auger Observatory consists of 1600 
Surface Detector (SD) nodes spread over an area of around 
3000 $\mathrm{km}^2$.  Similarly, LOFAR (Low Frequency ARray) 
for radio astronomy contains 8000 sensors and is spread over an 
area of more than 100 km in diameter. All these systems belong to
the class of spatial sensor networks.

A characteristic feature of many of these spatial sensor networks is that 
each sensor node in the network has a local $``$trigger'' and not 
only collects data but also sends it to a central unit for further analysis.  
However, as the amount of data that needs to be processed grows, and the area 
becomes larger, realizing the communication path from sensor to central
unit becomes problematic. Firstly, the bandwidth requirements can be met 
only with special resources like fibre optic links. 
Installing a fibre optic or any other wired infrastructure is often infeasible 
(certainly in large areas). Consequently, wireless communication remains 
the only viable option. However, wireless communication has its own drawbacks,
notably, limited bandwidth and unreliability of communication.  
Secondly, the central unit can easily become a bottleneck when the \emph{number} 
of nodes grow and the \emph{rate} at which data is collected grows.

Considering that sensed data often contains lots of raw measurements that 
will eventually be aggregated into much more informative units requiring 
much less space, local processing by nodes is essential for scalability.
In many cases, significant improvements can be made if nodes in each other's 
proximity collaborate in the data analysis. This increases the 
complexity (computing power, storage capabilities and costs) of a single 
node 
and how to balance these issues is the main topic of this paper. 
Collaborative local data analysis may result in sending truly relevant 
aggregated (and location based) information to a central unit for further 
analysis, and may be the only path toward scalable solutions.

Local data analysis imposes additional requirements. For example, there may be 
a need for additional communication with neighboring nodes to reach a decision, 
leading to relatively high bandwidth usage. 
Likewise, while awaiting results from neighbors, temporary storage requirements 
may be fairly high unless special measures are taken. Finally, we need to take 
into account that local analysis may be computationally so demanding that 
algorithms may need to be tuned to what sensor nodes can realistically accomplish.

As it turns out, there are many tradeoffs to consider in building scalable 
solutions based on collaborative local data analysis.  
For example, how often and when should nodes exchange information?  
A low frequency of data exchange may lead to highly bursty traffic that 
may exceed local bandwidth constraints and that incur local buffering demands 
--- which can be resolved by increasing the frequency of exchanges.  
However, frequent exchanges incur much higher energy costs caused by communication.

In this paper, we consider a specific, challenging application to illustrate 
the exploration of the design space for collaborative local data analysis in 
spatial sensor networks. The application involves the detection of ultra-high 
energy cosmic rays using a wireless sensor network.  
The application demands high communication bandwidth and involves large amounts 
of in-network data processing. On the other hand, the sensor nodes are resource 
constrained in terms of energy budget and capacity regarding computation and storage.
In~\cite{ded:12}, we presented a distributed event detection algorithm which is 
entirely based on collaborative local data analysis. We explored the application-level 
resource requirements such as communication bandwidth for a certain level of performance 
in an unreliable communication environment. 

We make two contributions. First, in contrast to the work described 
in~\cite{ded:12}, we explore in this paper the spectrum of tradeoffs 
that need to be considered while building scalable solutions based on 
collaborative local data analysis. 
The application that we consider is a natural fit for collaborative 
local data analysis. Therefore, a distributed system concept to detect 
high-energy cosmic-rays was proposed in~\cite{Clay1992101}. 
However, the concept was not studied in further details.  
This is the first paper to our knowledge to explore the possibility of 
applying collaborative local data analysis in large-scale spatial wireless 
sensor networks to detect ultra-high energy cosmic rays. 
Second, we provide all the necessary details required to come to full 
understanding of our distributed event detection algorithm. 

The rest of this paper is organized as follows.  
Section~\ref{Sec:SysMod} presents our system model, specifying the assumptions 
made and semantics of events continuously occurring in our system.  
In Section~\ref{sec:algo} we describe our distributed event detection 
algorithm in detail. 
We illustrate, in Section~\ref{Sec:DesignSpaceAnalysis}, the tradeoffs to 
consider while building scalable solutions based on collaborative local 
data analysis. Section~\ref{Sec:Method} describes the experimental setup, performance 
metrics of our algorithm, and methodology to compare centralized and distributed event 
detection algorithms. In Section~\ref{Sec:Results} we provide an experimental 
evaluation of our proposed distributed algorithm based on simulations. 
We discuss the centralized event detection approach and compare it with 
our distributed local detection algorithm in terms of bandwidth requirement in 
Section~\ref{Sec:Tech}, thereby quantifying the scalability and efficiency 
of our distributed approach. In Section~\ref{Sec:RelWork} we highlight related work.
Finally, in Section \ref{Sec:Con}, we conclude the discussion and present 
our future work.

 \section{The System Model}
\label{Sec:SysMod}

Cosmic rays are high-energy charged particles, \textit{e.g.} protons or
atomic nuclei, that may originate from astrophysical objects such as
supernova remnants and active galactic nuclei.  Despite over a century of study,
however, their origin and acceleration process is still under debate.  

At the highest energies, the flux of cosmic rays decreases to 1 particle per
square kilometer per century, making direct detection infeasible.  Instead,
large spatial areas are instrumented with particle detectors that
then measure the \emph{extensive air shower} created when the cosmic ray
interacts with the atmosphere of the Earth. In this work, we refer to the
spatial area hit by a cosmic-ray air shower as the \emph{event region}.

Multiple techniques can be used to detect cosmic-ray air showers.
The Pierre Auger Observatory in Argentina uses a hybrid array of
water-Cherenkov particle detectors and fluorescence telescopes to record
the shower and indirectly measure the direction, energy, and composition of
the original cosmic-ray~\cite{Abraham:2004dt}. Additionally, an enhancement 
is being deployed to detect the 
radio emission from the air showers: AERA~\cite{VLVNT:12}, the Auger Engineering Radio
Array. Because of the demands of the radio-detection method in particular
(see Section \ref{Sect:Motivation}), AERA is particularly suited for
consideration as a testbed for collaborative local data analysis.  We 
describe the AERA antenna array, and its triggering scheme, as a model for
these studies, while noting that the distributed event-detection scheme can
also be applied to other large-scale cosmic-ray air-shower experiments.

\subsection{System Setup}

We consider a vast field covered by a large collection of antenna stations.  Each antenna station -- from now
on called a \textbf{station} is a wireless sensor. The station can sense radio signals (in a specific frequency range)
and can communicate with neighboring stations in the field through a low-power wireless medium. Each station
is attached with a standalone energy-harvesting device (e.g., a solar panel), implying a modest energy budget
per station. Each station has limited processing capabilities and a storage capacity in the order of a few
hundred megabytes. The latter may seem much, but the incoming stream of raw, digitized samples that need to be
analysed for cosmic rays is such that storage is quickly filled up. As a consequence, data analysis needs to
be done within a limited time in order to free storage space.  

We assume that a GPS receiver is attached to each station allowing (1)~the clocks of stations to be globally
synchronized,\footnote{The accuracy is maintained within 10 nanoseconds through special devices} and
(2)~stations to be aware of their location. The stations are assumed to be stationary.  Each station is
capable of communicating with at least one other station in the field. Furthermore, the system has been set up
in such a way that under normal conditions the network of sensor nodes is strongly connected. In this paper, to
illustrate the essentials of our solution, we assume that all the communication channels are reliable.

Each station relays its data to a base station called the Central Radio Station (CRS). The CRS has
comparatively high capacity and sufficient energy.  Among different possibilities of placement of stations in
the field are a \emph{grid-based} placement (e.g, triangular, rectangular, or hexagonal etc.)  and a
\emph{uniform random} placement. We assume a grid-based placement.

\subsection{neighborhood Semantics}

We consider two notions of neighborhood of a station:

\begin{enumerate}

\item \textbf{Geographical neighborhood:} The geographical neighborhood of a station is defined as the set of stations within a distance $D$ from the station.  We assume that a station has knowledge about its
  geographical neighbors. The geographical neighborhood of a station $s$ may change due to addition,
  removal, or failure of one or more stations within distance $D$ from $s$. In this paper, we assume there is
  no change in geographical neighborhood.

\item \textbf{Network-level neighborhood:} In this case, the set of stations is represented by a graph
  $G(V,E)$, where vertex set $V$ represents the set of all stations and set $E$ contains an edge $(u,v)$ if
  $u$ can directly communicate with $v$. The network-level neighborhood of a station $s$ is then defined as
  the set of stations within $N$ hops from the station $s$. We assume direct communication only between
  geographical neighbors.

\end{enumerate}

\subsection{Event Semantics}

As described in~\cite{VLVNT:12} each station picks up radio signals with an antenna. These signals are digitized and filtered locally. The
filtered signal is analysed for pulses above a certain threshold, which may indicate the occurrence of a
cosmic ray. Such an event generates what is called an \textbf{N1 trigger}. 
In fact, the N1 trigger is equivalent to what is called the $``$level 2" trigger in~\cite{VLVNT:12}. 
Each trigger is timestamped at nanoseconds resolution. For each trigger, in addition to the timestamp, a digitized portion of the signal of 12.5 kilobytes is also buffered at the station. This buffered data is called \textbf{event data}. The event data along with the timestamp is sent to the CRS upon positive decision through a data analysis procedure; otherwise both the timestamp and event data are ignored.

The triggers of two geographically neighboring stations are said to be \emph{coincident} if their timestamp difference $\Delta T$ is less than $T_{c}$, the travel time of light in a straight line from one station to the other station. $T_{c}$ is also called coincidence window. An N1 trigger at a station promotes to an \textbf{N3 trigger} in two cases: 
\begin{itemize}
\item[(1)] the N1 trigger at a station is coincident with N1 triggers of at least two other geographical
  neighbors
\item[(2)] the N1 trigger at a station is coincident with an N3 trigger of any of its geographical neighbors.
\end{itemize}

Note that an N3 trigger fulfils the minimal requirement for what is called a Level 3 
trigger in~\cite{VLVNT:12} where the CRS forms the Level 3 by computing coincidence among multiple 
  (at least three) adjacent stations.

Normally, an N1 trigger at a station is discarded if it does not promote to an N3 trigger. 
However, due to link failures, it may not be possible to safely discard data, in which case 
the N1 trigger is still sent to the CRS.

\begin{figure}[htp]
   \centering\framebox{\includegraphics[scale=0.95]{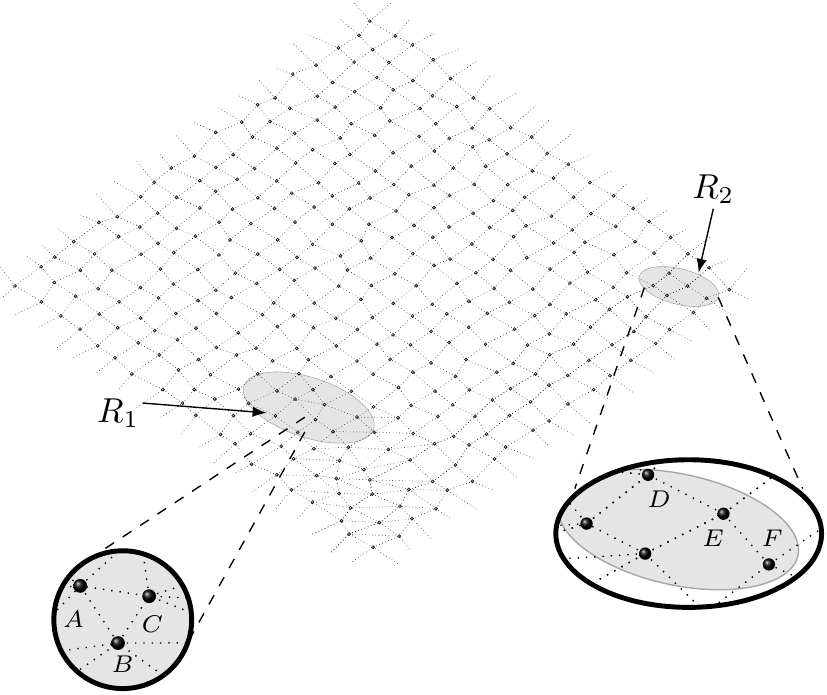}}
\caption{Two independent cosmic-ray air shower events. The event regions 
are shown shaded. Stations $A$, $B$, and $C$ illustrate a general case in 
which their N1 triggers promote to N3 triggers. Stations $D$, $E$, and $F$ 
illustrate a special case of promoting the N1 trigger of station $F$ to an 
N3 trigger. Note that station $F$ has only one neighbor in the event region.}
\label{Fig:AirShower}
\end {figure}

Figure~\ref{Fig:AirShower} illustrates the occurrence of two independent cosmic ray air shower events. The
event regions are shown shaded and labelled as $R_1$ and $R_2$. Each station in the event region has an N1
trigger. For illustration, we assume that the triggers in an event region are coincident. The assumption holds
for both $R_1$ and $R_2$. We have the following two cases:

\begin{itemize}

\item \textbf{General case:} Each of the stations $A$, $B$, and $C$ or any 
other station in $R1$ can be in coincidence with at least two other 
geographical neighbors in $R1$ and promotes its N1 trigger to an N3 trigger.

\item \textbf{Special Case:} Station F, as opposed to stations $D$ and $E$ or 
any other station in $R_2$ has only one geographical neighbor in the event 
region. Apparently, the N1 trigger of station $F$ cannot find coincidence with 
two other neighbors in $R_2$.  However, station $F$ is required to promote its 
N1 trigger to an N3 trigger. Any cosmic-ray detection technique must be able to 
handle this special case besides the general case.

\end{itemize}

\begin{figure}[htp]
  \centering\framebox{\includegraphics[scale=0.30]{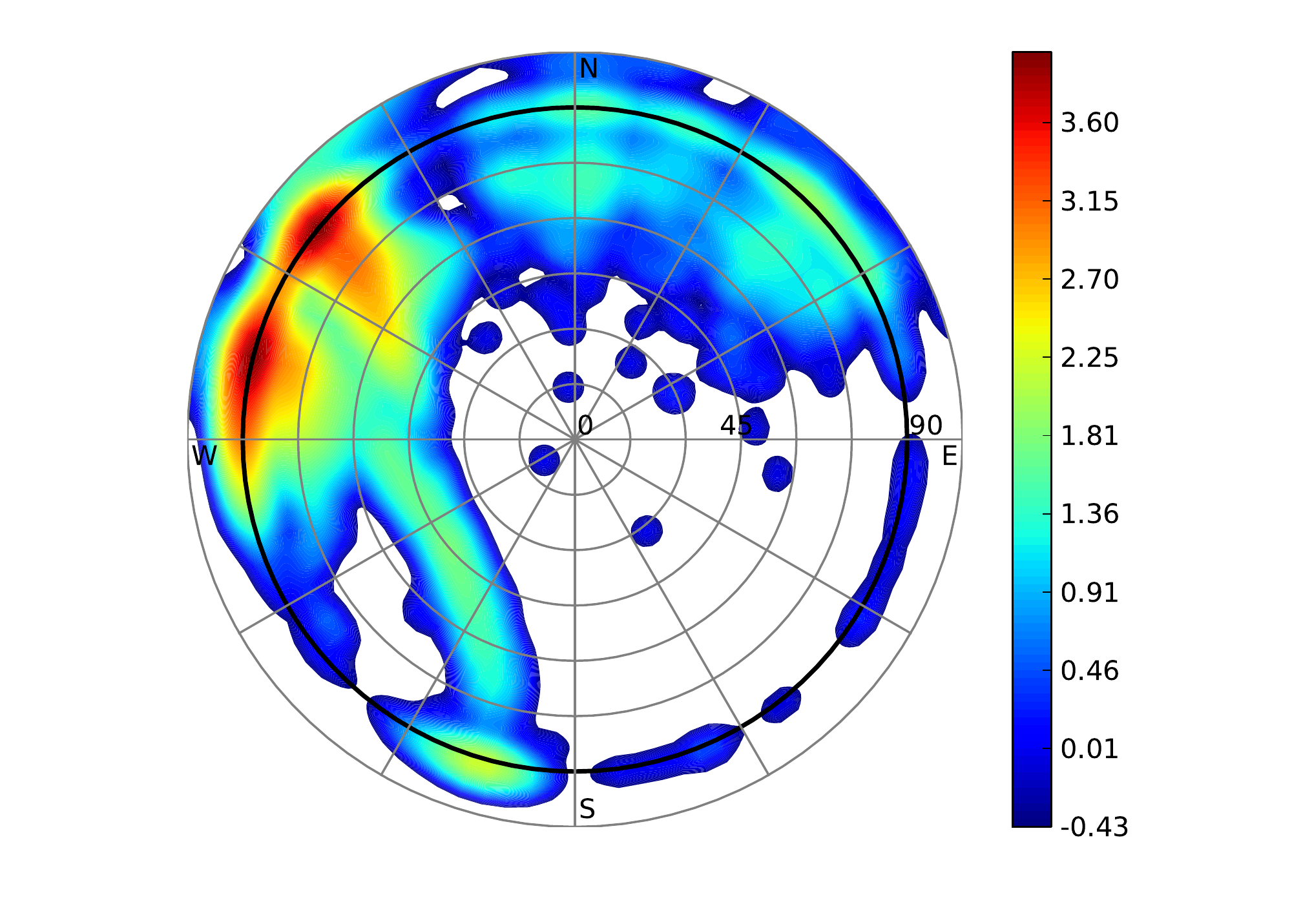}}
  \caption{A polar skymap of the reconstructed direction of sample Level 3 triggers,
    showing several man-made pulse radio sources on the horizon. The color scale
    indicates $\log(\mathrm{event density / a.u.})$. 
    The figure is \revision{reproduced from}~\cite{VLVNT:12}.}
  \label{Fig:SkyPlot}
\end {figure}

Figure~\ref{Fig:SkyPlot} is a sample skymap of Level 3 triggers measured at the CRS~\cite{VLVNT:12}. 
It shows the sky in polar coordinates, where the center is directly overhead, and the dark circle near the edge is the horizon. 
The $z$-axis (color scale) is the Level 3 trigger density in 
$\log(\mathrm{event density / a.u.})$.
It is clear from Figure~\ref{Fig:SkyPlot} that the dominant number of Level 3 triggers 
are caused by several point sources from the horizon. There are relatively fewer 
Level 3 triggers that indicate the direction of arrival of cosmic rays. 
This implies the need for criteria to filter out Level 3 triggers that are caused by 
point sources from the horizon. To carry out such filtering, we introduce a sense 
of direction.

The direction of the signal that caused the Level 3 trigger is reconstructed 
using timestamps and geographical positions of the stations that took 
part in the coincidence. Note that we assume direction reconstruction only for N3 triggers. 
The reason for considering only N3 triggers is explained later in Section~\ref{Sect:Motivation}. 
The direction reconstruction uses what is known 
as a \emph{plane wave fit}. The reconstructed direction is represented as 
a tuple of \emph{zenith} and \emph{azimuth} angles. An N3 trigger becomes 
an \emph{event of interest} if the direction of signal causing the N3 trigger 
is within a user-defined range of directions representing the horizon. Note, however, that 
the range representing the horizon is crucial to the performance of the algorithm. A small range 
will allow more N3 triggers to be reported to the CRS, whereas a large range may lead to producing 
\emph{false negatives}: discarded N3 triggers that were caused by an actual cosmic ray air shower.

\begin{figure}[htp]
\centering
\framebox{\includegraphics[scale=0.65]{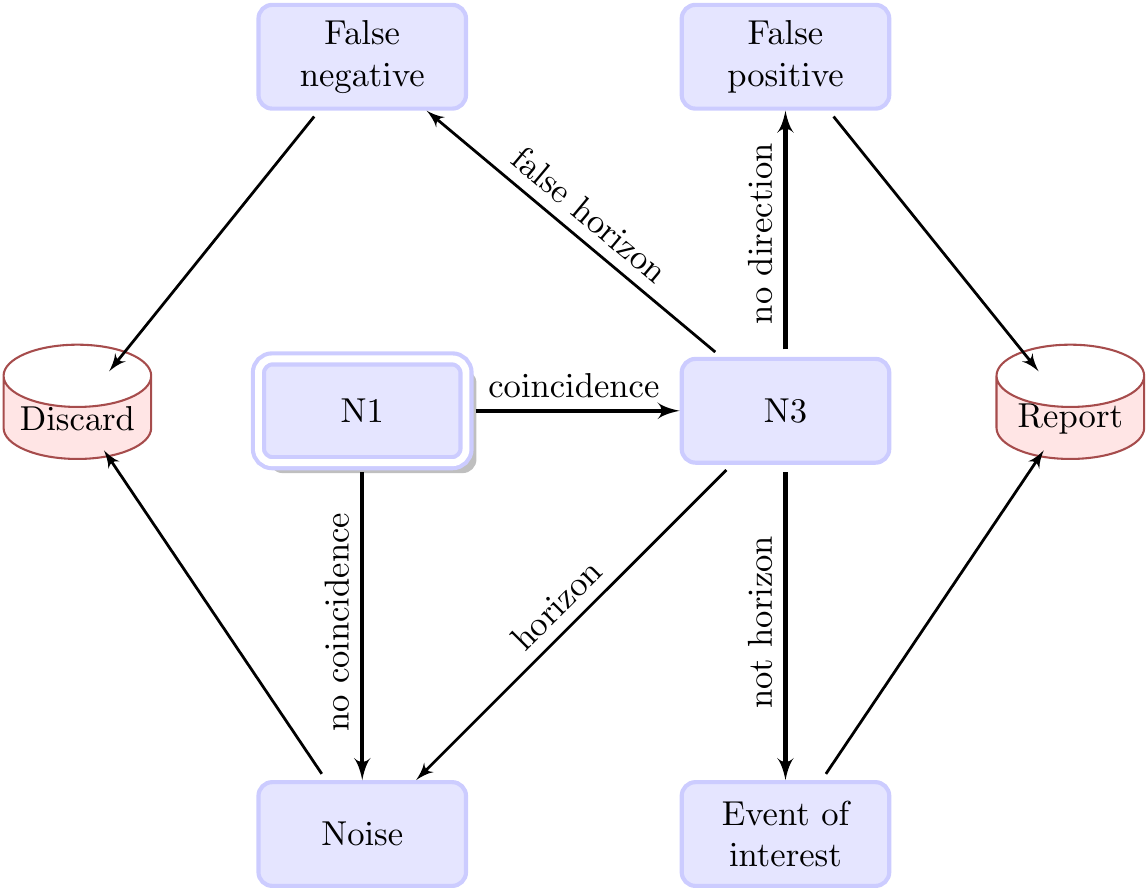}}
\caption{State transition diagram of N1 trigger.}\label{Fig:EventsFlow}
\end{figure}

The direction reconstruction process may fail for various reasons. First, the 
timestamps of N3 triggers may not be all from the same (real) signal. 
For example, 1 or 2 timestamps may be from accidental coincidences, and in this 
case there is no unique direction. Second, the timestamps are all from a real signal, 
but considering that the direction reconstruction process uses heuristics to 
compute direction, it may not converge and compute an incorrect result.
In either case of direction reconstruction failure, the N3 trigger is considered as a \emph{false positive}. In case of a false positive, the timestamp and associated event data of the N3 trigger is sent to the CRS. Figure~\ref{Fig:EventsFlow} summarizes all possible state transitions of an N1 trigger.

\subsection{\label{Sect:Motivation}Motivation for Collaborative Local Data Analysis}

A challenge of the radio detection of cosmic rays is that the antennas sense 
not only radio pulses emitted from air showers but are triggered also by pulsed 
random noise and man-made disturbances (e.g., power transformers or
airplanes).  While several techniques have been developed to distinguish
air shower pulses from man-made noise (see Ref.~\cite{VLVNT:12}),
the system must still be robust to elevated and highly variable trigger rates.

A typical initial trigger rate at an AERA antenna is roughly 200 Hz.
Since each trigger has an associated event data, this means that every
detector must, in principle, relay a huge amount of data to the CRS over
wireless links, which is practically impossible.  This is the primary
motivation for a hierarchical trigger scheme.  Requiring at least 3
antennas in time coincidence can reduce the rate to around 20 Hz.

As shown in Figure~\ref{Fig:SkyPlot}, around 90\% of these
triggers can be localized at the horizon, and are therefore likely man-made
noise.  However, a directional reconstruction requires timing information
from multiple antennas.  In a centralized system, this decision
must be made at a single point that collects all the trigger times from all
of the stations.  

Collaborative local data analysis can play a pivotal role to discard the
uninteresting man-made noise events. In principle, each station is able to
find a time coincidence with its 
geographical neighbors, if it knows their timestamps.  A station can then
decide locally whether its \textbf{N1} trigger is an event of interest. This idea is
the main motivation behind devising a distributed event detection algorithm
that will allow to build scalable and energy efficient solutions for
ultra-high-energy cosmic-ray detection. 

\section{The Distributed Event Detection Algorithm}
\label{sec:algo}

The goal of our Distributed Event Detection (DED) algorithm is twofold.  Firstly, to decide whether an N1
trigger generated locally at a station should be promoted to an N3 trigger. Secondly, provided that an N1
trigger has been promoted to an N3 trigger, to reconstruct, locally, the direction of the radio signal that caused
the trigger. Recall that the direction reconstruction helps filter out man-made disturbances including signals
transmitted by point sources from the horizon.  The direction is reconstructed using timestamps and positions
of the neighboring stations that helped to promote the N1 trigger to an N3 trigger.

\subsection{Algorithm I: A Conceptual View of DED}
\label{subsec:algo1}

As a first attempt, we design a distributed algorithm that takes decisions based on local information. A
station can communicate with all of its geographical neighbors, and we assume no communication failure.

The basic idea of the distributed event detection is the following. Whenever an N1 trigger occurs at a
station, the station stores the N1 trigger locally and informs all of its geographical neighbors. 
The information consists of a message of type~\type{N1Entry} whose structure is shown in Figure~\ref{fig:datatypes1}.
Furthermore, when a station receives N1 triggers from its neighbors, it looks for a coincidence of the
received triggers with its local ones. A station promotes its N1 trigger to an N3 trigger if its N1 trigger
has coincidence with N1 triggers of at least two geographical neighbors.

To cover stations on the boundary of an event region with only one geographical neighbor in the event region,
a station not only requires to broadcast its N1 triggers, but also its N3 triggers. We call the latter type of broadcast messages \textbf{advertisements}. 
The information contained in an advertisement are of type~\type{AdvertEntry} 
whose structure is shown in Figure~\ref{fig:datatypes1}. An \type{AdvertEntry} 
entry contains (1)~the timestamp of the local N1 trigger that was promoted to 
an N3 trigger, and (2)~the two N1 triggers that had coincidence with the local 
N1 trigger.
An advertisement helps a station promote its N1 trigger to N3 trigger if its N1 trigger 
has coincidence with the N3 trigger contained in the advertisement message.

\begin{figure}[!htb]
\begin{minipage}[t]{0.45\linewidth}
{\small
\begin{alltt}
    \cdef{type} N1Entry = \cdef{record}
         srcId : \ccon{integer};
         seconds : \ccon{integer};
         nanoSeconds : \ccon{integer};
    \cdef{end};\ccom{\{ type N1Entry \}}
\end{alltt}
}
\end{minipage}
\begin{minipage}[t]{0.45\linewidth}
{\small
\begin{alltt}
    \cdef{type} AdvertEntry = \cdef{record}
         srcId : \ccon{integer};
         seconds : \ccon{integer};
         nanoSeconds : \ccon{integer};
         neighborN1 : \ccon{N1Entry[]};
    \cdef{end};\ccom{\{ type AdvertEntry \}}
\end{alltt}
}
\end{minipage}
\caption{Data types for Algorithm I.}
\label{fig:datatypes1}
\end{figure}

\begin{figure}[!htb]
\begin{center}
{\small
\begin{alltt}
\textbf{/*** On Local N1 Trigger ***/} 
\ccom{// Runs when a local N1 trigger occurs at station \(p\)}
\copr{localCache.add}(\(N1(p)\)))  
\(\ccod{\forallt}  q\;\in\;Neigh\sb{p}  \ccod{\doo}\) 
  \msnd{N1(p)}{q}\emptyline\emptyline
\textbf{/*** Receive from neighbor ***/}
\ccom{// Runs when receiving an N1 trigger or Advert}
\mrcv{N1}{q}
\(\ccod{\orr}\)
\mrcv{advert}{q}\emptyline\emptyline
\(\ccod{\foranyt} q\sb{i},\;q\sb{j}\;\in\;Neigh\sb{p} \ccod{\doo}\) 
  \ccod{if} \copr{coincidence}\((N1(p),\;N1(q\sb{i}),\;N1(q\sb{j}))\) \ccod{then}
     \(N1(p)\;\goes\;N3(p)\)
     \(process(N3(p))\)
     \(\ccod{\forallt}  q\;\in\;Neigh\sb{p}  \ccod{\doo}\) 
       \msnd{advert(p)}{q}
     \copr{localCache.remove}(\(N1(p)\)))\emptyline\emptyline
\(\ccod{\foranyt} q\;\in\;Neigh\sb{p} \ccod{\doo}\)
  \ccod{if} \copr{coincidence}\((N1(p),\;advert(q))\) \ccod{then}
     \(N1(p)\;\goes\;N3(p)\)
     \(process(N3(p))\)
     \copr{localCache.remove}(\(N1(p)\)))\emptyline\emptyline
\textbf{/*** On Remove Trigger ***/} 
\ccom{// Runs when a local N1 trigger is marked for} 
\ccom{// removal under the cache eviction policy}
\ccod{if} \copr{NOT} \copr{isDecided}(\(N1(p)\)) \ccod{then}
  \copr{apply user defined criteria}
\end{alltt}
}
\end{center}
\caption{Pseudocode for Algorithm I.}
\label{fig:alg1}
\end{figure}
%

Figure~\ref{fig:alg1} shows the pseudocode of our algorithm. When an N1 trigger 
occurs at a station $p$, it adds the trigger to its local cache. 
The local cache has a limited capacity. On the other hand, there is a continuous 
stream of new triggers arriving into the cache due to which the cache may become 
full thus requiring a \emph{cache eviction} policy. To this end, we consider the 
cache as a FIFO queue and remove the oldest trigger when the cache becomes full. 
Therefore, the algorithm is required to process each trigger before it is 
removed under the cache eviction policy. This requirement imposes a time 
constraint on the processing of each trigger. Obviously, the time span within which 
a trigger should be processed by the algorithm depends on the size of the cache.
Next, the station also informs all of its geographical neighbors about 
the occurrence of its N1 trigger by broadcasting a message of type \type{N1Entry}.  

The station then starts listening to (1)~N1 triggers from any of its geographical neighbors, and
(2)~advertisement messages from any of its geographical neighbors.  Whenever an N1 trigger is promoted to an
N3 trigger, it will be discarded by the station after a computation that involves the following:

\begin{itemize}
\item Direction reconstruction of the signal that caused the trigger.
\item Deciding, based on the reconstructed direction, if the trigger is an \emph{event of interest}, \emph{false positive}, or \emph{noise}.
\end{itemize}

\noindent
In case that station $p$ promotes its N1 trigger to an N3 trigger because of coincidence with N1 triggers of
two of its geographical neighbors, the station informs all its geographical neighbors by broadcasting a
message of type \type{AdvertEntry}. 
On the other hand, if station $p$ promotes its N1 trigger to an N3 trigger because of coincidence
with an N3 trigger of any of its geographical neighbors, the promoted N3 trigger is not broadcast to the
neighbors. The reason for this is that station $p$ has only one neighbor in the event region which has
already promoted its corresponding N1 trigger to N3. For station $p$, broadcasting the advertisement in this
case is useless.  Once an N1 trigger is promoted to an N3 trigger it is removed from storage, in order to be
sent to the CRS.

When a trigger is marked for removal under the cache eviction policy and it has not yet been decided by the algorithm (if the trigger is an N3) then 
there are two possibilities to decide about this trigger before being removed.  First, under the 
assumption of reliable communication, the trigger is discarded; concluding 
it is not an N3 trigger.  Second, under the assumption of unreliable communication, 
which we do not consider in this paper, station $p$ may apply some user-defined policy. 
The user-defined policy is usually a heuristic filter that is applied to the local N1 
triggers of a station. One such heuristic filter is that the N1 triggers occur at 
repeated time intervals.  For example, some N1 triggers are caused by
electrical power lines passing over the field where antenna stations are installed.  These triggers show
periodicity correlating with the usual AC power line frequency of 50 or 60 Hz and should be ignored.

\subsection{Algorithm II: A Periodic Broadcast Algorithm for DED}
\label{subsec:algo2}

We notice that each station broadcasts its local N1 triggers at a rate of 200 Hz next to broadcasting its
advertisement messages. Our previous algorithm suffers from high bandwidth consumption. To reduce bandwidth
consumption, we use an alternative algorithm discussed in this section.

The idea behind this algorithm is that we can save a significant amount of bandwidth by grouping local N1
triggers that share the \type{seconds} field. The triggers are bundled such that N1 triggers with different
values of \type{nanoseconds} field share a common value of the \type{seconds} field.

\begin{figure}[!ht]
\centering\framebox{\includegraphics[scale=0.65]{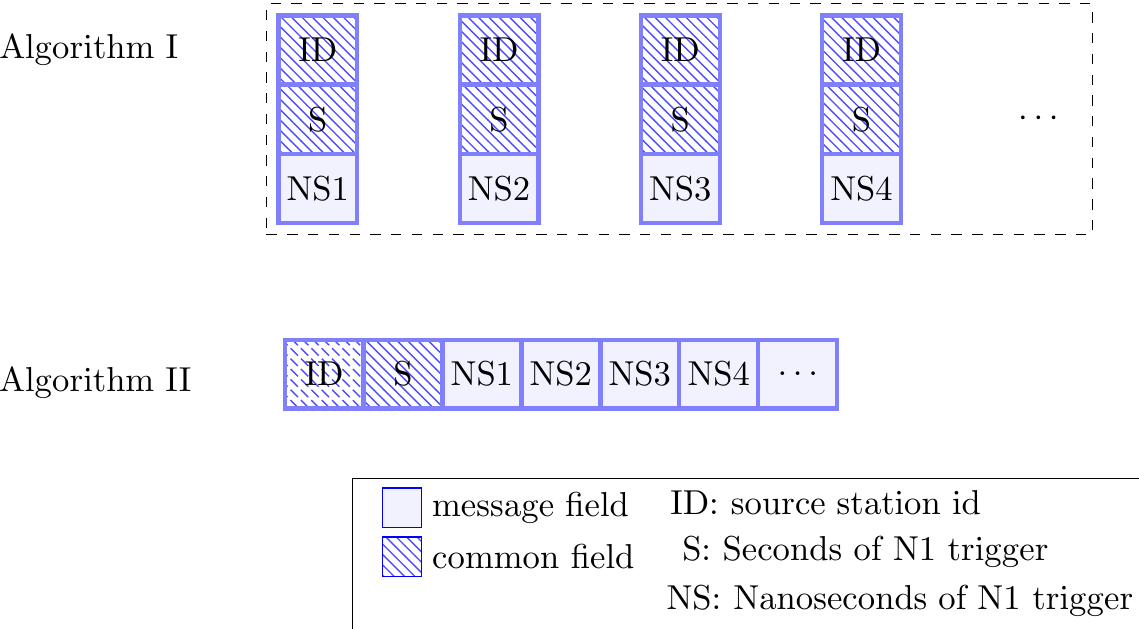}}
  \caption{Grouping N1 triggers having the same value of $\tt{seconds}$ field into a bundle.}
  \label{Fig:DeltaCompression}
\end{figure}

Figure~\ref{Fig:DeltaCompression} compares messages used in both algorithms to broadcast four N1 triggers
generated at a station to its geographical neighbors. We assume that the timestamps of all the four triggers
have the same value for the \type{seconds} field. It is obvious that our first algorithm uses one message per
N1 trigger and broadcasts an extensive amount of redundant information in the form of the \type{srcId} and the
\type{seconds} fields. Our modified algorithm can save a significant amount of bandwidth, without loss of
accuracy, by simply grouping together the N1 triggers having the same value for the \type{seconds} field. We
call the N1 triggers grouped in this way an \textbf{N1 bundle}, as shown in Figure~\ref{fig:datatypes_2}.  In
our system, N1 triggers occur at each station at an average rate of 200Hz.  We see that by \emph{bundling}
these N1 triggers together, having the same value for the \type{srcId} field and assuming they share the same
value for the \type{seconds} field, the bundle will carry only one instance of both \type{srcId} and
\type{seconds} fields and a list of distinct \type{nanoseconds}. This reduces, in principle, the bandwidth
consumption by around 66\%.

Similar to the N1 bundle formation, the local N3 triggers that share the same value for the \type{seconds}
field can be bundled as an \textbf{advertisement bundle}, also shown in Figure~\ref{fig:datatypes_2}.  The
bandwidth consumption can further be reduced by compressing bundles before their broadcast. Note, however,
that this involves computational overhead caused by compression and decompression.

\begin{figure}[!htb]
\begin{minipage}[t]{0.45\linewidth}
{\small
\begin{alltt}
    \cdef{type} N1Bundle = \cdef{record}
         srcId : \ccon{integer};
         seconds : \ccon{integer};
         nanoSeconds : \ccon{integer[]};
    \cdef{end};\ccom{\{ type N1Bundle \}}\emptyline\emptyline
    \cdef{type} N1Entry = \cdef{record}
         srcId : \ccon{integer};
         seconds : \ccon{integer};
         nanoSeconds : \ccon{integer};
    \cdef{end};\ccom{\{ type N1Entry \}}\emptyline\emptyline
\end{alltt}}
\end{minipage}
\begin{minipage}[t]{0.45\linewidth}
{\small
\begin{alltt}
    \cdef{type} AdvertBundle = \cdef{record}
         srcId : \ccon{integer};
         seconds : \ccon{integer};
         advertBundleEntry  
                 : \ccon{AdvertBundleEntry[]};
    \cdef{end};\ccom{\{ type AdvertBundle \}}
    
    \cdef{type} AdvertBundleEntry = \cdef{record}
         nanoSeconds : \ccon{integer};
         neighborN1 : \ccon{N1Entry[]};
    \cdef{end};\ccom{\{ type AdvertBundleEntry \}}\emptyline\emptyline
\end{alltt}
}
\end{minipage}
\caption{Data types for Algorithm II.}
\label{fig:datatypes_2}
\end{figure}
\begin{figure}[!htb]
\begin{center}
{\small
\begin{alltt}
\textbf{/*** On Local N1 Trigger ***/} 
\ccom{// Runs when a local N1 trigger occurs at station \(p\)}
\copr{localCache.add}(\(N1(p)\)))  
\copr{N1Bundle.add}(\(N1(p)\)))\emptyline\emptyline
\textbf{/*** Active thread ***/} 
\ccom{// Runs every T seconds}
\(\ccod{\forallt}  q\;\in\;Neigh\sb{p}  \ccod{\doo}\) 
  \msnd{N1Bundle(p)}{q}
\(\ccod{\forallt}  q\;\in\;Neigh\sb{p}  \ccod{\doo}\)
  \msnd{AdvertBundle(p)}{q}\emptyline\emptyline 
\textbf{/*** Passive thread ***/}
\ccom{// Runs when receiving an N1Bundle or AdvertBundle}
\mrcv{N1Bundle}{q}
\(\ccod{\orr}\)
\mrcv{advertBundle}{q}\emptyline\emptyline
\(\ccod{\forallt} entry\;\in\;Bundle \ccod{\doo}\)
  \copr{coincidence}\;\(=\)\;\copr{localCache.coincidence}(\(entry\)))
  \ccod{if} \copr{coincidence} \ccod{then}
     \(N1(p)\;\goes\;N3(p)\)
     \(process(N3(p))\)
     \(\ccod{if}  entry\;\in\;N1Bundle  \ccod{\doo}\) 
       \copr{N1Bundle.add}(\(entry\)))
     \copr{localCache.remove}(\(N1(p)\)))\emptyline\emptyline
\textbf{/*** On Remove Trigger ***/} 
\ccom{// Runs when a local N1 trigger is marked for} 
\ccom{// removal under the cache eviction policy}
\ccod{if} \copr{NOT} \copr{isDecided}(\(N1(p)\)) \ccod{then}
  \copr{apply user defined criteria}
\end{alltt}
}
\end{center}
\caption{Pseudocode for Algorithm II.}
\label{fig:alg2}
\end{figure}

Figure~\ref{fig:alg2} shows the pseudocode of our revised algorithm. 
When an N1 trigger occurs at a station $p$, it adds the trigger to its 
local cache. Due to limited storage capacity each trigger is discarded from
the local cache under a certain \emph{cache eviction} policy. We use a simple 
cache eviction policy that removes the oldest trigger from the cache. 
The trigger is also added to a local N1 bundle that will be broadcast to 
geographical neighbors of the station.

The algorithm executes two threads: active and passive.  The active thread is executed periodically. It
broadcasts N1 bundles and advertisement bundles of a station to the geographical neighbors of the station.
The passive thread listens to incoming messages. Upon receipt of an N1 bundle or an advertisement bundle from
a geographical neighbor, the thread looks for a coincidence of each trigger in the bundle with the local N1
triggers.  Whenever an N1 trigger is promoted to an N3 trigger, it will be discarded if it came from an
invalid direction as before.

If a coincidence has been found among the local N1 trigger and N1 triggers of the geographical neighbors,
then these N1 triggers are added to the advertisement bundle.

When a trigger is marked for removal under the cache eviction policy and it has not yet been
decided by the algorithm (if the trigger is an N3) then it 
is treated exactly the same way as before.
\section{Design Space Analysis}
\label{Sec:DesignSpaceAnalysis}

We measure the accuracy of our algorithms according to the number of false negatives and efficiency according to the number of false positives and the amount of communication among geographical neighbors during event detection. 
The algorithm is said to have produced a \textbf{false negative} if a trigger is not reported to the CRS which would have been marked by a reference/ideal algorithm as part of a potential cosmic-ray air shower. 
Similarly, the algorithm produces a \textbf{false positive} if a trigger is reported to the CRS because there is not sufficient information to conclude that it is caused by man-made noise and hence discard it.
Ideally, the algorithms should neither produce false negatives, nor false positives. However, due to the constraints of limited amount of communication bandwidth, storage, and computational power, we need to consider various tradeoffs. These tradeoffs may cause the algorithms to produce false negatives and false positives.

\begin{figure*}[th!]
\begin{center}
 \includegraphics[scale=0.95]{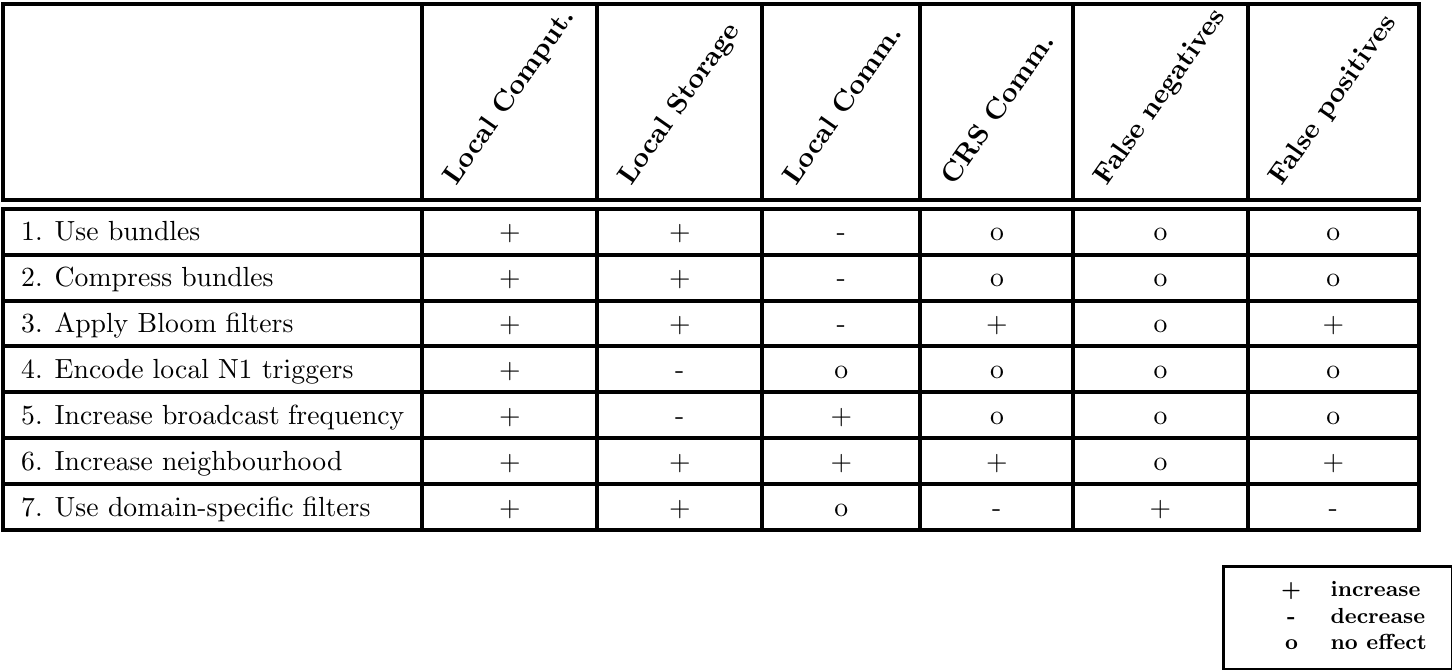} 
\caption{Design space exploration for collaborative local data analysis techniques.}
\end{center}
\label{Fig:tradeoffs}
\end{figure*}

We use our proposed first algorithm under ideal conditions as reference point for comparison.  There are
several cases related to our revised algorithm where tradeoffs can be considered. These cases are discussed
below and summarized in Figure~\ref{Fig:tradeoffs}.

\begin{enumerate}

\item By using bundles, we reduce local communication at the price of increased local computation and
  (temporary) storage.

\item Compressing bundles will further decrease communication costs, yet also increase local computational effort. In addition, more local storage is needed in comparison to not using any bundles at all, although compression will help to keep this required additional storage low.

\item Probabilistic data structures like Bloom filters~\cite{Bloom:1970} for exchanging N1 triggers, result in reduced communication during local data analysis. Actually, each local N1 trigger is hashed 
to a single bit and instead of broadcasting a bundle of local N1 triggers, a bit vector representing the triggers is broadcast to the neighbors. Then a node compares its local bit vector with the bit vectors received from the neighbors. 
The lookup operation is cheaper. Moreover, being a characteristic of the Bloom filters, no 
false negatives are produced. However, this technique increases the number of false positives, 
in turn, implying increased communication overhead with the CRS.

\item We can reduce the memory usage at the cost of increased computation as follows. 
Instead of storing the timestamp of a local N1 trigger in a pair of \type{seconds} and 
\type{nanoseconds} fields, we store it as an offset to a certain base time. 
The full information on an N1 trigger can be computed back using this base time and the offset.

\item If the frequency with which a station broadcasts its N1 bundle and advertisement bundle 
is increased, then it will allow stations to decide on their local N1 triggers earlier. Since 
an N1 trigger is removed from the local cache when it has been evaluated, increasing the 
broadcast frequency will result in reduction of memory usage for storing N1 triggers. 
On the other hand, the increased broadcast frequency will increase the number of bundles 
exchanged overall, and this increased number of bundles will increase the computational 
overhead required for each bundle. Moreover, an increased broadcast frequency will also 
consume more bandwidth which means consuming more energy.

\item For the same transmission range, increasing the neighborhood size beyond a certain minimum has a
  negative effect on performance. The reason is that more N1 triggers are generated whose collaborative
  processing consumes resources for no gain. Note that this statement is based on the assumption of reliable
  communication channels. In case of unreliable communication channels, increased neighborhood size will
  improve the robustness of the system.

\item As mentioned earlier, a station may apply some user-defined heuristic filter to the local N1 triggers of
  the station.  The heuristic filter discards those N1 triggers which fall within certain time windows along
  the time domain. If the length of the window is kept too small then fewer N1 triggers will be discarded. On
  the other hand, keeping a larger window the filter may discard some N1 triggers which are potential N3
  triggers. This situation gives rise to false negatives.

\end{enumerate}

\section{Experimental Setup and Methodology}
\label{Sec:Method}

We carried out simulations to demonstrate the accuracy and efficiency of our distributed event detection
algorithms. The simulations were conducted using the OMNET++~\cite{omnet} simulation environment. The OMNET++ platform is expressive, efficient, modular, and prevailing as the de-facto simulation environment for mobile ad-hoc and sensor networks~\cite{weingartner2009performance}.
We used traces of N1 triggers collected from AERA testbed. The current testbed consists of 24 stations
and uses wired infrastructure for communication between stations and the CRS.  The data analysis procedure in
the testbed is centralized: every station sends its N1 triggers to the CRS for analysis. 
It is important to emphasize that the occurrence of N1
triggers is independent of the data analysis procedure. So the N1 triggers generated in the testbed with wired
infrastructure and centralized data analysis procedure could still be used to validate and analyse our
proposed solution based on wireless infrastructure and collaborative local data analysis procedure.

\subsection{Validation}
\label{Sec:validation}
To validate our approach, we compared the functionality of our algorithms based on distributed event detection (DED) to that of the 
centralized event detection (CED). 
In the CED, each station sends timestamps of its N1 triggers to the CRS. The CRS performs multi-station coincidences. If three or more 
stations are in coincidence the corresponding group of triggers is promoted to Level 3. The CRS then requests each of these stations to send their corresponding event data.

The current implementation of CED disregards geographical neighborhood for the computation of Level 3 triggers. This implies that the 
triggers from every two stations in the system are considered for possible coincidence. In this case the coincidence window is kept as 
large as the travel time of light across the entire array (not just the distance between the two stations which are checked for coincidence). The array width is approximately 
750 meters for AERA. Thus, an \emph{array-level coincidence window} $T_{c} =2.5 \mu s$ is used in the CED.

There are several implications of the coincidence criteria based on the array-level $T_{c}$. First, although an exhaustive search for pairwise coincidence 
is fast for a small array like the current AERA, it may be computationally infeasible as the array size grows to thousands 
of stations. Second, there is an upper bound (approximately $10Km^{2}$) on the surface area that can be hit by a highest-energy cosmic-ray 
air-shower. Therefore, searching for coincidence between two stations that are far apart from each other (beyond the mentioned upper bound) 
is meaningless.

On the other hand, DED assumes that a station can communicate only with its geographical neighbors, and thereby search for coincidence 
in the geographical neighborhood. Moreover, the coincidence criteria is based on the travel-time of light between the pair of stations whose 
triggers are checked for coincidence. Currently, the average interstation distance in AERA is $150m$, which means $T_{c}=0.5\mu s$. 
Comparing the array-level $T_{c}$ of CED and pair-level $T_{c}$ of DED we observed that the set $S_{ced}$ of N1 triggers promoted to Level 3 
by CED is a superset of the corresponding set $S_{ded}$ produced by DED. This is indeed an expected outcome because DED uses more strict 
coincidence criteria than the CED. 

An important question that arises here is whether the difference of the sets of triggers $S_{ced} \backslash S_{ded}$ are actual false negatives. 
To answer this question, we inspect closely the CED system. In addition to the different coincidence window, CED employs a domain-specific filter 
called dynamic histogram~\cite{VLVNT:12}. This filter removes hot spots~(triggers which arrive periodically from the same direction) from the set of triggers $S_{ced}$ and results in a set $S_{hist}$, 
representing groups of triggers that are potential cosmic-ray air-showers. We observed that the 
difference $S_{hist} \backslash S_{ded}$ is empty, and hence, DED does not drop any trigger that is part of a potential cosmic-ray 
air-shower. However, to make a fair comparison between the performances of CED and DED, we imposed the same set of assumptions on CED and DED, 
namely, geographical adjacency (and, thus, the coincidence window $T_{c}=2.5\mu s$). 
For these two assumptions 
DED should detect the same set of Level 3 triggers as detected by CED. 
In other words, if $S_{ced} = S_{ded}$ then the output of DED is valid and the related performance comparisons will be fair. 

We impose geographical adjacency on the CED system as follows. 
Consider the network as a simple undirected graph $G(V,E)$, where 
stations are nodes in $V$. Two nodes are connected by an edge in $E$ 
if these nodes are geographical neighbors. 
All stations involved in a Level 3 trigger $i$ captured by CED form a subgraph $G_i(V_i,E_i) \in G$. 
(Note that a subgraph $G_i$ may be a disconnected graph). 
According to our \emph{connected component criteria}, two nodes in $G_i$ have an edge $e \in E$ between them, if they are geographical neighbors, and 
are in coincidence with each other. We search for connected components in $G_i$, and select all connected components 
(CC) of $G_i$ with the size $\geq 3$. 
The resulting set of CC for all Level 3 will satisfy geographical adjacency.

\subsection{Performance Metrics and System Parameters}
In order to analyse the effect of using collaborative local data analysis, we considered a representative
station \emph{S} in our simulation. Recall that our centralized approach would have sent all the N1 triggers
generated at station \emph{S} to the CRS. We were interested to see whether station \emph{S}, under our
collaborative local data analysis approach, can actually reduce the data that is sent to the CRS.  To that
end, we monitored the frequency of the following variables for station \emph{S}:
\begin{itemize}
\item N1 triggers
\item N3 triggers
\item False positives
\item False negatives
\item Events of interest
\end{itemize}

\noindent
Here, an event of interest is defined as the N3 trigger for which the direction reconstruction process is successful and the \emph{zenith} angle (in degrees) lies outside the interval [90-$\eta$, 90+$\eta$].   The signal arriving from an angle within the interval [90-$\eta$, 90+$\eta$] is considered as a signal from the horizon (\ie\ man-made disturbance). Here, $\eta$ is an adjustable parameter to analyse the tradeoff between accuracy and efficiency of direction reconstruction filter.

As argued in Section~\ref{subsec:algo2}, our enhanced algorithm can save a significant amount of bandwidth by
bundling consecutive local N1 triggers before broadcasting to geographical neighbors.  To see this effect we
compared the bandwidth utilization of our basic algorithm and our enhanced algorithm with respect to the
following variables for station \emph{S}:
\begin{itemize}
\item Transmission rate (bits/sec)
\item Reception rate (bits/sec)
\end{itemize}

Finally, we examined the effect of additional computation at the stations on communication.
To that end, we applied compression to the N1 bundles and advertisement bundles in our 
enhanced algorithm. We used lossless compression (\emph{zlib-1.2.5}) so that 
the receiving station is able to reconstruct the original bundle.

\begin{figure}[!t]
  \hspace*{-0.5cm}
   \begin{minipage}[t]{0.49\linewidth}
    \centering
  \includegraphics[scale=0.5]{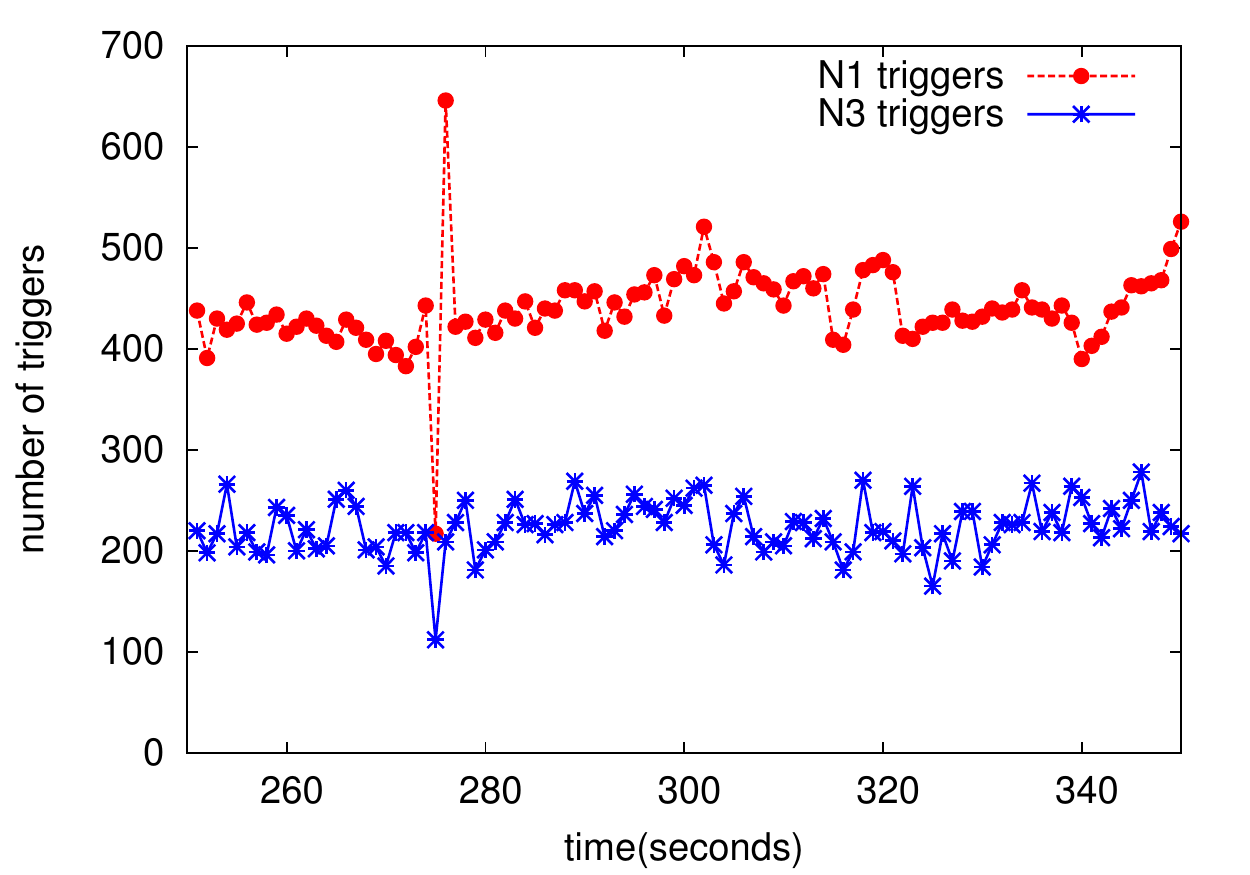}
  \caption{Collaborative local data analysis at station \emph{S}.}
  \label{Fig:DFilter}
  \end{minipage}
  \hspace{0.1cm}
  \begin{minipage}[t]{0.49\linewidth}
    \centering
  \includegraphics[scale=0.5]{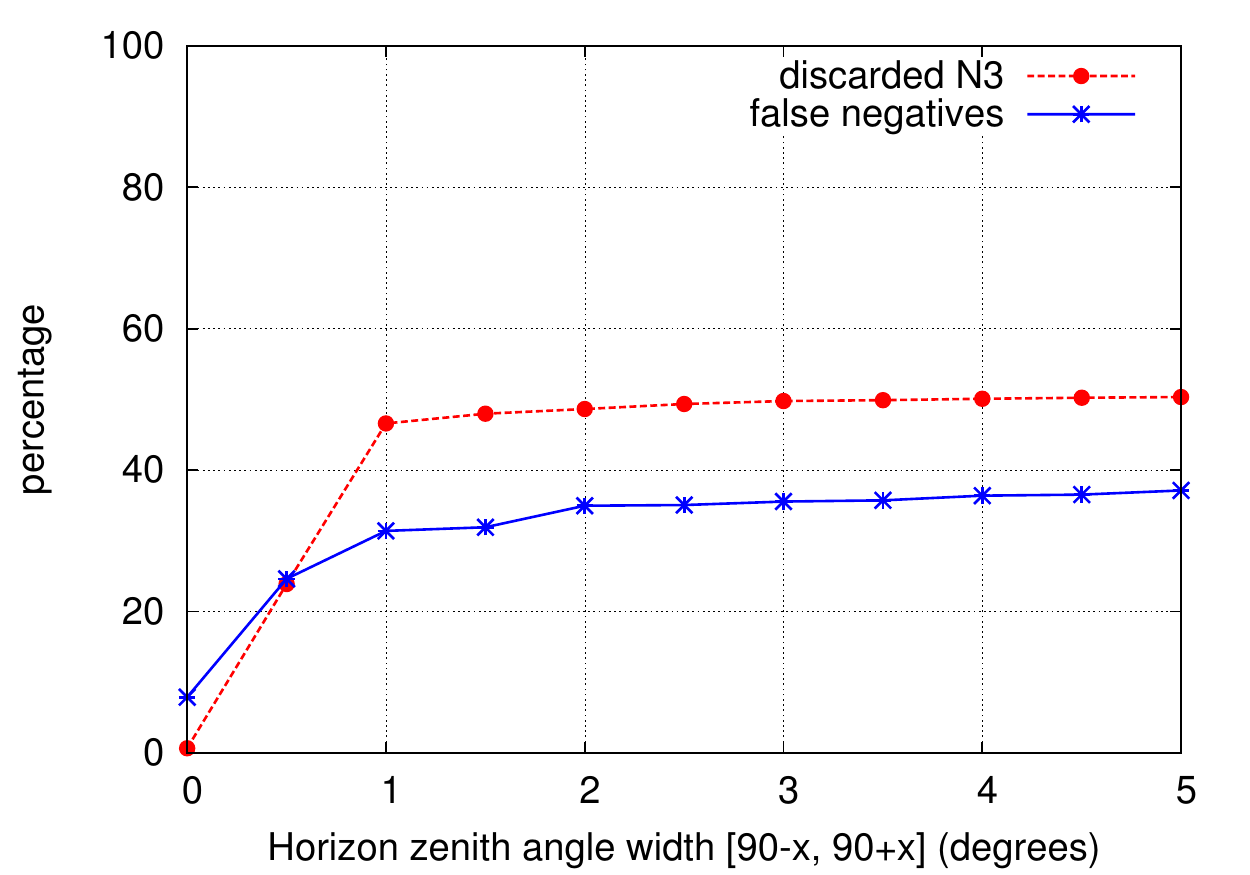}
  \caption{The tradeoff between discarding N3 triggers based on reconstructed direction at the cost of producing false negatives.}
  \label{Fig:FalseNeg}
    \end{minipage}
\end{figure}

\begin{figure}[!b]
  \hspace*{-0.4cm}
\subfigure[Average transmission rate]
{
  {\label{Tx_a1a2a2c_a}\includegraphics[scale=0.5]{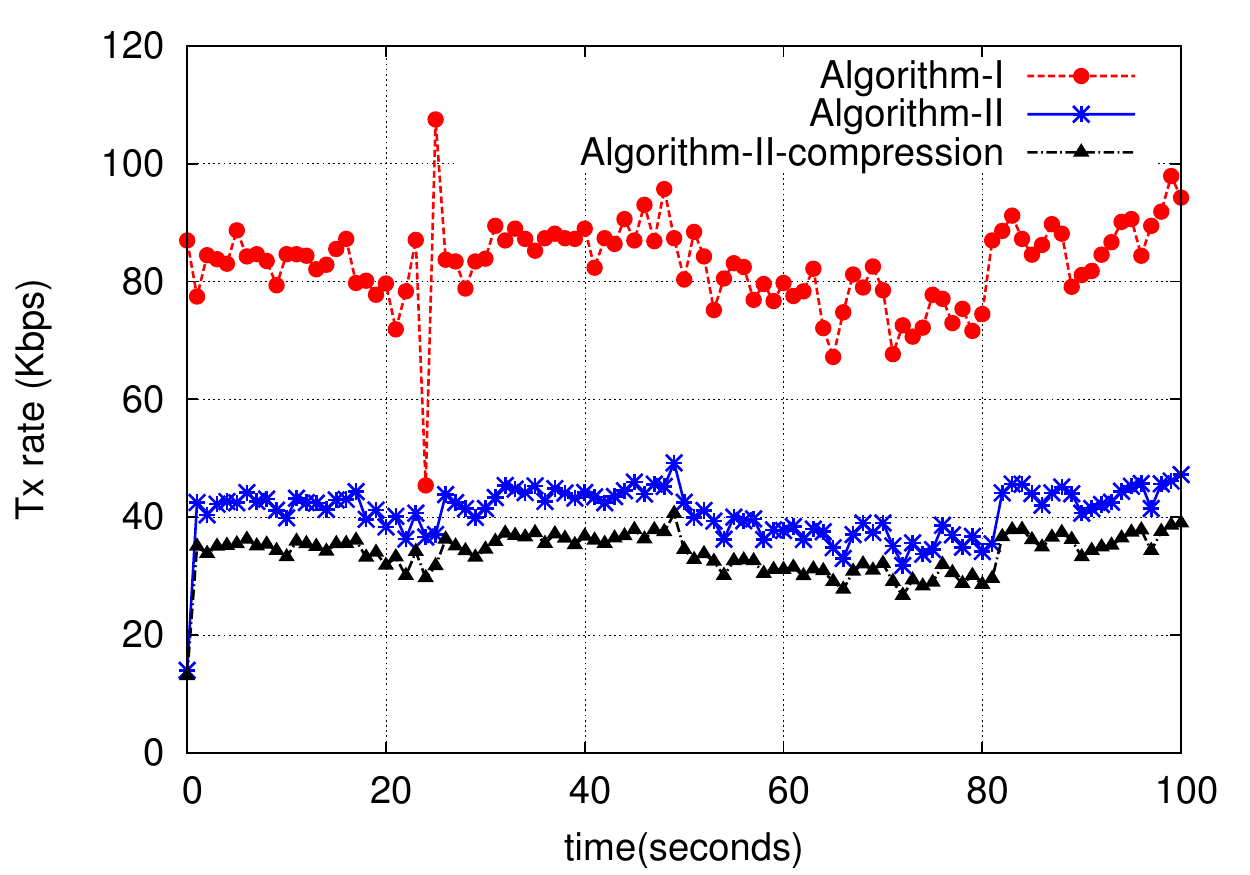}}
}
  \hspace*{-0.4cm}
\subfigure[Average reception rate]
{
  {\label{Tx_a1a2a2c_b}\includegraphics[scale=0.5]{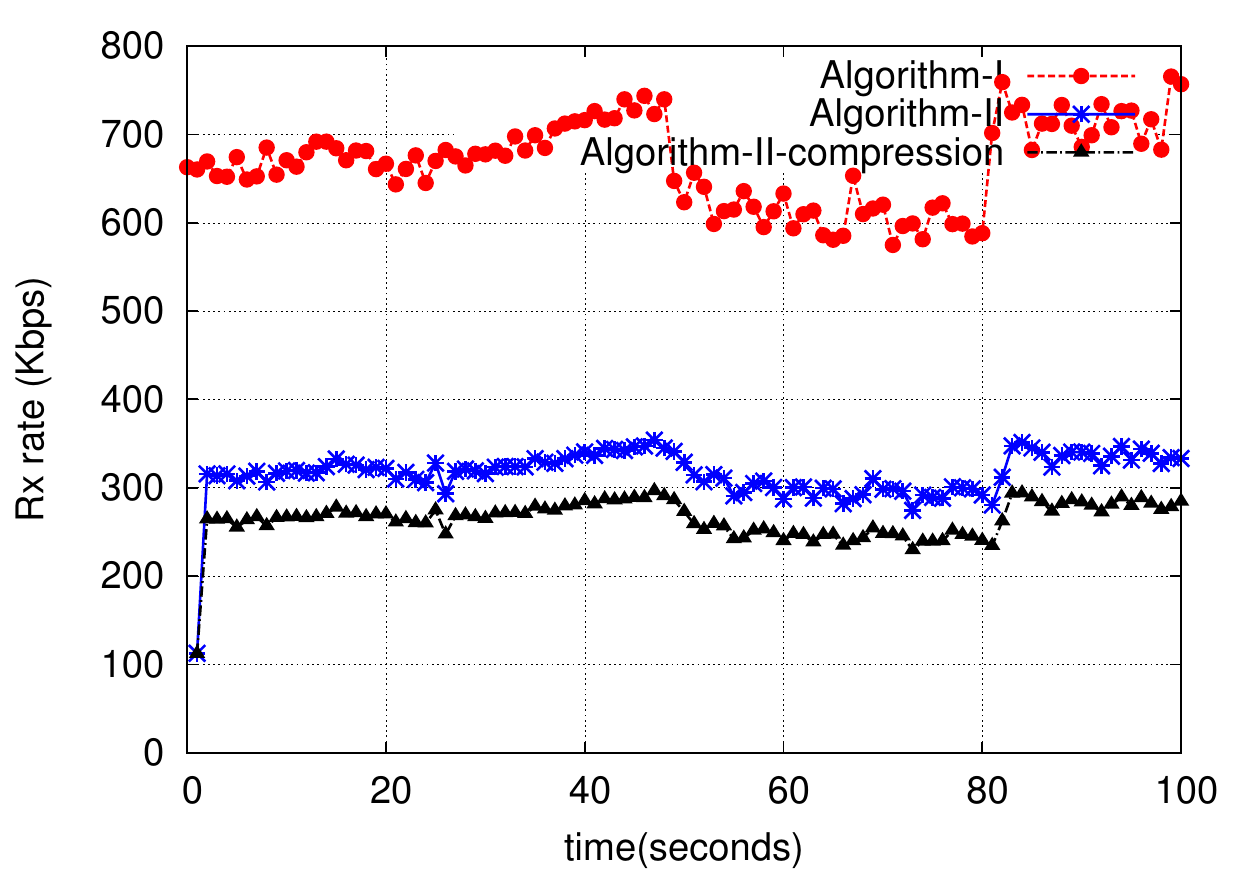}}
}
\caption{Performance comparison of Algorithm-I and Algorithm-II with respect to communication bandwidth consumption for station \emph{S}.}
\label{Fig:Comm_a1a2a2c}
\end{figure}
\section{Results}
\label{Sec:Results}
We first focused on validation of our collaborative local data analysis approach. 
We followed our validation method explained in Section~\ref{Sec:validation} and confirmed that 
our algorithms indeed observed the same N3 triggers that should have been noticed. 
This means that our algorithms are functioning correctly.

Let us now consider the local filtering capability of our algorithms, 
by focusing on our representative station \emph{S}. Figure~\ref{Fig:DFilter} 
shows the local data analysis performed at station \emph{S} over a period of 
100 seconds. 
The frequency of N1 triggers (generated at station \emph{S}) 
during the experiment is shown. Next, as a result of executing our 
distributed event detection algorithms, the fraction of N1 triggers 
at station \emph{S} that were promoted to N3 triggers is shown. 

Wireless communication offers limited bandwidth, therefore, efficient 
utilization of bandwidth becomes essential. From Figure~\ref{Fig:DFilter} 
we can see that our distributed algorithms enable a station to discard a 
huge amount of data by communicating only with its geographical neighbors. 
Only a relatively small amount of data is chosen to be sent to the CRS for 
further analysis. This is a positive indication for efficient bandwidth utilization.

Our algorithms may produce false positives. Therefore, we monitored the occurrence of false positives during the experiments. A \emph{false positive} is a consequence of the direction reconstruction process failure. As discussed in Section~\ref{Sec:SysMod}, there are two reasons 
for this failure.  First, there might be no real signal underlying the corresponding 
N3 trigger. If we were sure that there is no real signal underlying the N3
trigger, then instead of reporting it as a false positive we could confidently discard 
it locally. This would enable the station to send even less data to the CRS. Second, 
the direction reconstruction process, using heuristics, may fail to reconstruct direction 
for an N3 trigger caused by a real signal. In both cases we need more efficient methods 
for direction reconstruction which will help in further reduction of the amount of data
sent to the CRS. However, in our sample trace of the N1 triggers for station 
\emph{S}, we did not observe any false positives. 

Efficient utilization of bandwidth, under high frequency of N1 triggers, also involves several tradeoffs. As an example we consider our definition of \textbf{event of interest}, which we defined to be an N3 trigger whose zenith angle (in degrees) lies outside the interval [90-$\eta$, 90+$\eta$]. Note that the interval represents the direction of signals from the horizon. We took $\eta \in \lbrace 0,0.5,1, ..., 4.5, 5 \rbrace$ and analysed the impact of width-of-horizon on accuracy and efficiency of the algorithm. The results are shown in Figure~\ref{Fig:FalseNeg}.
As we increase this interval, more N3 triggers will be discarded. This means that the amount of data that is sent to the CRS is further reduced and subsequently more efficient bandwidth utilization. On the other hand, keeping a large interval to represent the horizon may lead to producing \emph{false negatives}.
This situation shows a tradeoff between \emph{accuracy} and \emph{communication cost}.

It is important to note that both of our algorithms produce the same output in terms of N1 triggers promoted
to N3 triggers, false positives produced, and N3 triggers declared to be events of interest.  The only
difference between the two algorithms is in bandwidth utilization.  The enhanced algorithm is expected to
utilize the bandwidth more efficiently than the basic algorithm.  Figure~\ref{Fig:Comm_a1a2a2c} shows a
comparison of bandwidth utilization by our basic algorithm, our enhanced algorithm, and our enhanced algorithm
with bundle compression. The measurements were performed at our representative station \emph{S}. 
We see that there is a substantial difference between the bandwidth utilization of the basic algorithm and the enhanced version.

Figure~\ref{Fig:Comm_a1a2a2c} also demonstrates the effect of compressing N1 bundles and advertisement bundles
in our enhanced algorithm before broadcasting to its geographical neighbors. The bundle compression
significantly reduces the number of transmitted bits per second. The effect of compression becomes more
pronounced in case of reception of compressed bundles. This was to be expected considering the fact that
station \emph{S} receives from multiple neighbors. We believe the effect of compression can be made more
significant by applying compression techniques that consider the nature of the data we are handling.

\section{Centralized versus Distributed Event Detection}
\label{Sec:Tech}

The motivation for using DED is twofold. 
Firstly, the DED saves on bandwidth by avoiding transmission of N1 
triggers to the central station (both DED and CED detect the same set 
of N3 triggers). Second, DED has the advantage of spreading the computational 
load over the entire array instead of a single central computer. 
In this section we elaborate on when DED becomes more cost-effective 
(in terms of bandwidth) than CED. We moreover discuss the technological aspects 
of distributed event detection.

\newcommand{\nodeband}{\lambda_{\textit{station}}}
\newcommand{\cedband}{\lambda_{\textit{ced}}}
\newcommand{\dedband}{\lambda_{\textit{ded}}}
\newcommand{\adband}{\lambda_{\textit{advert}}}
\newcommand{\nband}{\lambda_{N1}}
\newcommand{\netsize}{\mathcal{N}}

\subsection{Bandwidth Usage}

We compare the bandwidth requirement of DED (algorithm-II) to that of CED. In general, 
the detection mechanism requires bandwidth in two phases: 1) event detection, and 2) 
routing the event data. As shown in the previous section, DED and CED produce the 
same number of N3 triggers, meaning that both approaches have the same bandwidth 
requirement for the second phase. Thus, we compare bandwidth requirement only for 
the event detection and exclude \emph{event data} routing in both CED and DED. The threshold network 
size is derived, based on bandwidth requirements, for which DED is more cost-effective 
than CED.

First, we estimate how much data is produced by a station. Each station bundles 
its N1 triggers in time intervals of one second. The bundle consists of the 
station identity (16 bits), seconds of the N1 triggers (32 bits), 
and list of nanoseconds (each ns is 32 bits). The station-level bandwidth 
requirement $\nodeband$ (bits/sec) can thus be estimated in terms of a frequency 
of N1 triggers $f_{N1}$ by the following equation:
\begin{equation}
\nodeband = 32 \cdot f_{N1} + 32 + 16 = 32 \cdot (f_{N1} + 1.5)
\label{bw_station}
\end{equation}
For CED, $\nodeband$ is the amount of data that needs to be routed, over 
$h$ hops, to the CRS for analysis. Here $h$ depends on the location of a 
station w.r.t. the CRS. Thus, the minimum requirement for the bandwidth 
by CED can be derived as follows:
\begin{equation}
\cedband = 32 \cdot (f_{N1} + 1.5) \cdot h
\label{bw_ced}
\end{equation}
Likewise, we estimate the bandwidth required by a station in DED. In DED, not 
only N1 triggers are bundled by a station but also N3 triggers of the current 
round are gathered in a separate advertisement bundle. The N1 bundle consists 
of station identity (16 bits), seconds (32 bits), and nanoseconds 
(each ns is 32 bits). The amount of data produced by an N1 bundle 
(in a time interval of one second) is:
\begin{equation}
\nband = 32 \cdot (f_{N1} + 1.5)
\end{equation}
Similarly, the advertisement bundle consists of station identity (16 bits), 
seconds (32 bits), and a list of type \type{AdvertBundleEntry}. Each 
\type{AdvertBundleEntry} has nanoseconds (32 bits) of the local N3 trigger, 
identities of two neighbors ($2 \cdot 16$ bits), and nanoseconds of the N3 triggers 
of these neighbors ($2 \cdot 32$ bits). The amount of data produced by an 
advertisement bundle (in a time interval of one second) is:
\begin{equation}
\adband = (32 + 2 \cdot 16 + 2 \cdot 32)  \cdot f_{N3} + 32 + 16 = 32  \cdot (4 \cdot f_{N3} + 1.5)
\label{bw_advert}
\end{equation}
Here $f_{N3}$ is the frequency of N3 triggers. 
The total bandwidth requirement $\dedband$ for a station in DED is 
the sum $\nband +\adband$. 

We noticed during our experiments that the ratio of promoted N3 triggers to the total N1 triggers is approximately $0.6$. 
For simplicity, we assume that this ratio is constant over time and is the same for all stations.
Then, substituting $f_{N3} = 0.6 \cdot f_{N1}$ in Equation~\ref{bw_advert}, we get:
\begin{equation}
\dedband = 32 \cdot (3.4 \cdot f_{N1} + 3)
\label{bw_ded_observed}
\end{equation}
In general case, i.e. for any ratio of $f_{N3}$ and $f_{N1}$, 
Equation~\ref{bw_ded_observed} becomes   
\begin{equation}
\dedband = 32 \cdot (\beta \cdot f_{N1} + 4)
\label{bw_ded_general}
\end{equation}
From Equation~\ref{bw_ced} and \ref{bw_ded_general}, we see that $h$ 
and $\beta$ are the determining factors for bandwidth requirements of 
CED and DED, respectively. 
In other words, CED and DED have the same bandwidth requirement if 
the average number of hops $\overline{h}$ for each bundle to travel 
from the source station to the CRS is equal to $\beta$.
Clearly, CED requires less bandwidth than DED if $\overline{h} < \beta$.
However, for $\overline{h} > \beta$ DED outperforms the CED. For our network 
configuration, triangular grid topology, and assuming that the CRS is located 
at the center of the network, the corresponding network size $\netsize$ can be 
determined by:
\begin{align}
\netsize = 1 + 6 \sum_{i=1}^{2 \overline{h} - 1} i, & &\text{where } \overline{h}=floor(\beta)
\label{nw_size}
\end{align}
(The central position of the CRS in CED provides the optimal communication path.) 
For the specific case of Equation~\ref{bw_ded_observed}, where $\beta=3.4$, the threshold network size $\netsize$ on or above which DED outperforms CED is 91. 

\begin{figure}[b!]
  \hspace*{-0.6cm}
\subfigure[Small-scale networks]
{
  {\label{bw_quan_small}\includegraphics[scale=0.5]{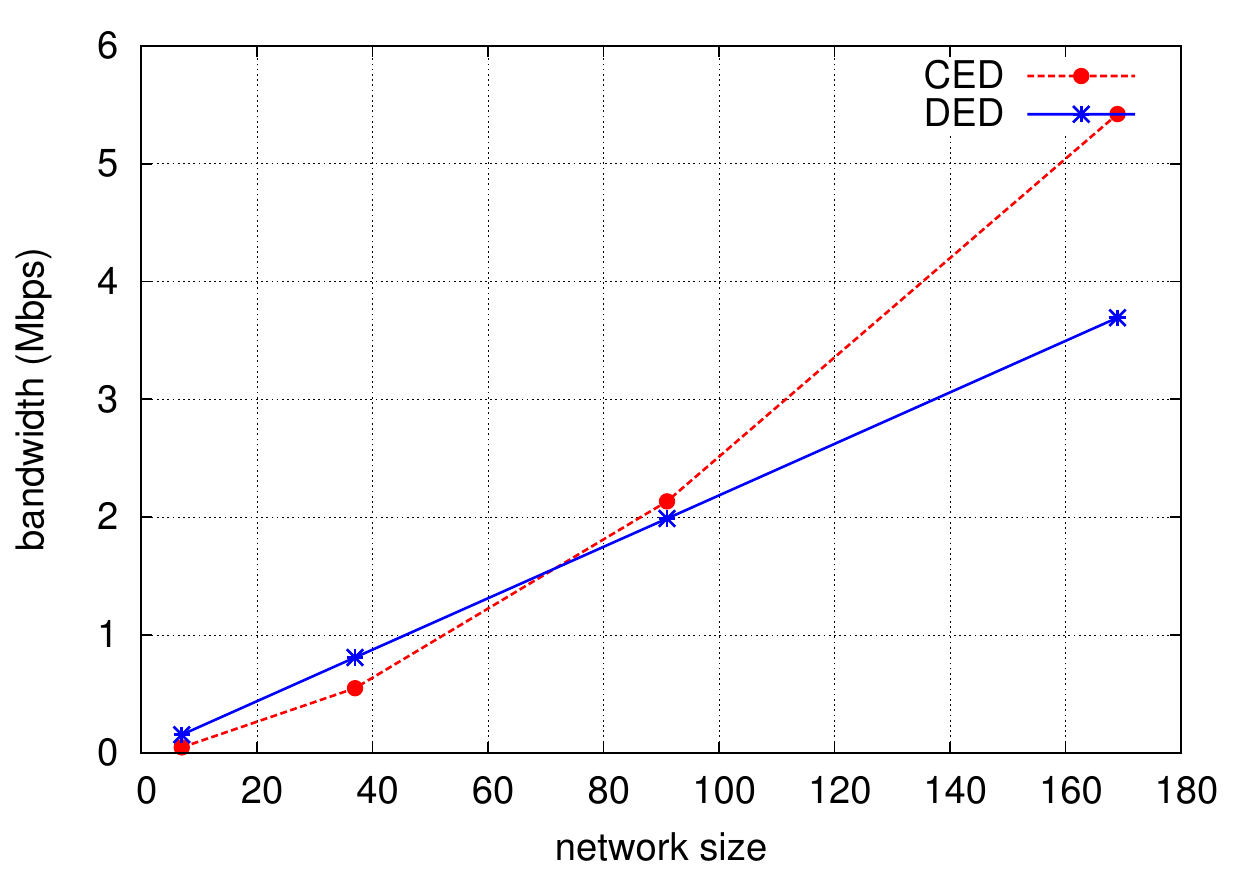}}
}
\subfigure[Large-scale networks]
{
  {\label{bw_quan_large}\includegraphics[scale=0.5]{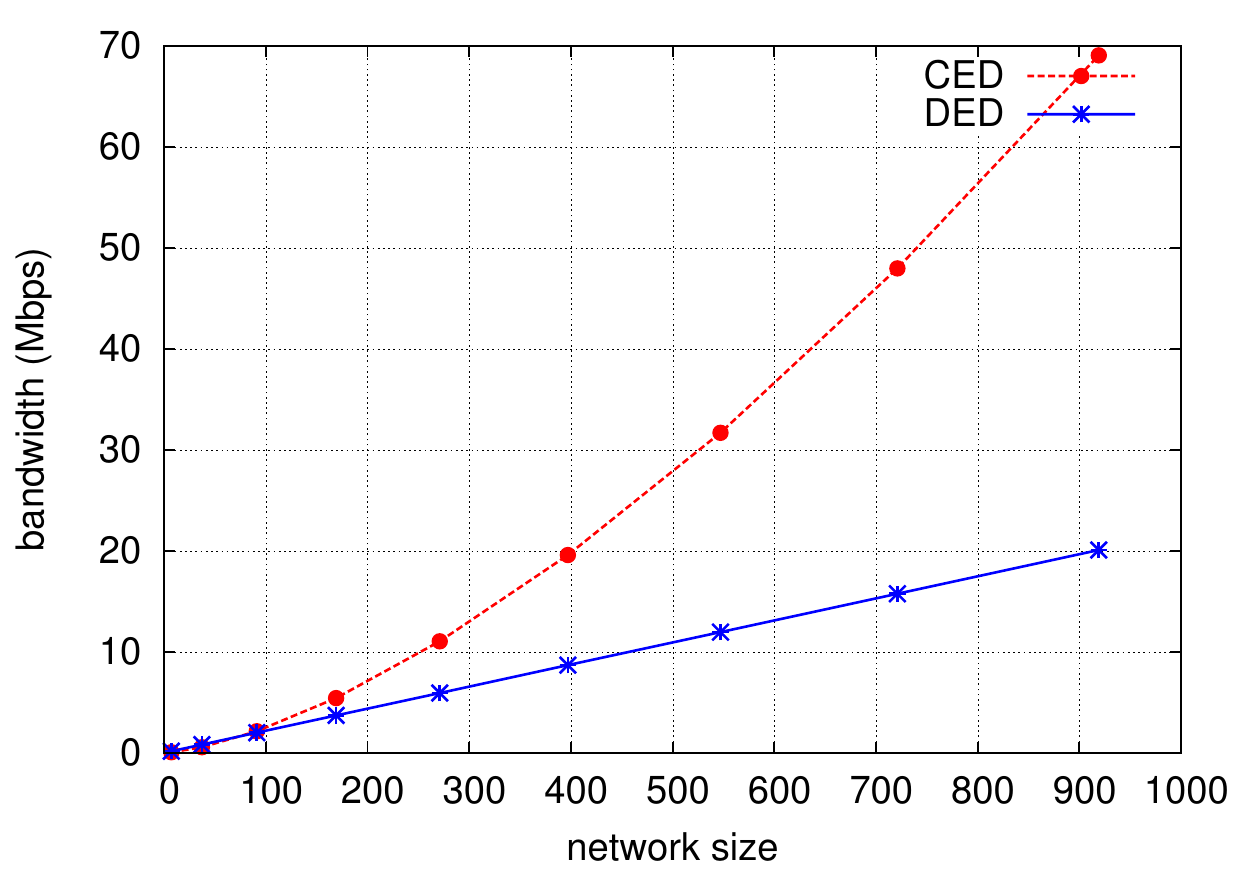}}
}
\caption{Bandwidth usage for CED and DED (during event detection phase). The x-axis represents size of a network. The y-axis represents the total amount of data transmitted in the network per second (during the detection phase). It is assumed that each station in the network has a trigger rate of 200~Hz.}
\label{Fig:BW_quantification}
\end{figure}

Figure~\ref{Fig:BW_quantification} elaborates further the comparison of bandwidth requirement for CED and DED for various network sizes. A network of size $\netsize$ is assumed to have a triangular grid topology. Furthermore, the stations are placed along the grid such that they form concentric hexagons. The CRS is in the centre. This means that the stations placed along the inner-most hexagon will be one hop from the CRS. Similarly, stations along the outermost $N^{th}$ hexagon will be $N$ hops away from the CRS. 

We see in Figure~\ref{bw_quan_small} that for smaller network sizes such as $\netsize \le 40$, CED has less bandwidth requirements. However, as the network size increases, the bandwidth requirement of CED becomes higher than DED. This can be seen for the threshold network size $\netsize = 91$. 
Figure~\ref{bw_quan_large} depicts a typical situation for the Auger North with 4400 SD stations and five concentrators. Every concentrator would have to cover approximately 900 SD stations. In this case the bandwidth requirement for CED will be $3.44$ times higher than DED.
\subsection{Technical Considerations}
The design of the trigger system for the Pierre Auger Observatory evolved over 
time to fulfil various objectives. These include scientific significance, 
quantity, and quality of data. The observatory was envisioned to have two 
sites, namely, Auger North and Auger South. 
The objective was to cover both the northern and southern hemisphere of the Earth. 
Moreover, to collect a large quantity of data, large geographical areas were 
instrumented on both sites. Currently, the AERA is under construction with 
the aim of getting high quality data that will help answer some questions 
related to high-energy cosmic-rays~\cite{astra-7-207-2011}. 

The changes in trigger systems reflected both in the design of detector 
stations and the related infrastructure for detector-to-CRS communication. 
These changes were made to overcome the then existing limitations of the 
trigger systems. However, they also posed new challenges.
We focus on these changes and discuss how the decisions taken may be 
helpful in the design of a distributed trigger for AERA to overcome the new 
challenges and exploit, at the same time, the opportunities in AERA for 
richer data collection.

The southern site, Auger South, uses wireless communication system organized 
in a two-layer hierarchy~\cite{Abraham:2004dt}. The area instrumented with 
detectors is divided into sectors. Each station communicates directly 
via directional antennas with a dedicated base station for that sector. 
Note that all the base stations are mounted on four concentrator towers.
The detector-to-base station communication uses custom radio hardware and 
proprietary network access protocols. 
The base stations are connected via a microwave network~(with a standard 
telecommunication architecture). The data at the base stations are transferred 
via the microwave network to the Central Data Acquisition System. 
The microwave network provides sufficient capacity including a margin for future use. However, the bottom layer involved in 
detector-to-base station communication was designed to support a bandwidth of {1200~bits/sec}.

The communication paradigm was reviewed for the northern site, Auger North, 
due to various reasons~\cite{CED_SD_North_Kieckhafer:2011zz}. These include 
increased bandwidth requirement for detector-to-base station communication, 
difference in terrain of both sites, and the cost of towers to cover the 
comparatively larger area of the Auger North. 
The detector array is considered as a wireless sensor network (WSN). Essentially, 
it is a peer-to-peeer (P-2-P) communication paradigm. More specifically, each detector 
can communicate with its geographical neighbors. A detector is enabled to 
send its data to the concentrator via multi-hop communication using neighboring 
detectors as intermediate relays.
Each detector is equipped with semi-custom radio hardware which uses 
four licensed, dedicated channels in the 4.6~GHz band. The WAHREN protocol~\cite{CED_SD_North_Kieckhafer:2011zz}, essentially a TDMA 
MAC with specific schedule for the Auger North setup, is implemented on 
top of the specified hardware.
Each detector requires the minimal bandwidth of 2400~bits/sec. 
Since detectors also relay data of their neighbors, the total bandwidth 
requirement is met at achievable bit rates, such as 11~Mbps. 
An important lesson learnt is that as the scale of required bandwidth per 
detector and the area instrumented with detectors increases, multi-hop 
communication is more cost-effective than single-hop detector-to-concentrator communication.

On the other hand, AERA antenna stations use different detection technique than 
the one used by Surface Detectors in Auger South and Auger North. Although the 
antenna stations collect richer data they also raise the detection-phase 
bandwidth requirement per antenna station (\textasciitilde 6900~bits/sec). 
Note that this bandwidth estimate is based on a typical trigger rate of 200~Hz. 
The trigger rate may be as high as between 400~Hz to 600~Hz. 
Based on the lesson learnt from Auger North and the highest bandwidth 
requirements of the antenna stations, localized event detection is a 
natural further step to cope with the high demands for bandwidth by 
antenna stations. 

A full design of communication system for DED is beyond the scope of this 
paper and requires further investigation. Nevertheless, we indicate two 
options that can be considered. First, investigate the TDMA schedule of 
the Auger North for a potential reconfiguration that allows each antenna 
station to exchange messages with its geographical neighbors. 
Second, the use of Commercial Off The Shelf (COTS) communication hardware. 
A potential candidate is ZigBee-Pro (IEEE-802.15.4, with operating frequency 
of 2.4~GHz in the ISM band). 
ZigBee-Pro offers an outdoor communication range up to 1500 meters and a 
data-rate upto 250Kbps. However, the MAC protocol used is based on CSMA. 
This means that packet collisions can be expected which will require 
redundant packet transmission. Therefore, further investigation is needed 
to experimentally evaluate the performance of DED executed on top of 
ZigBee-Pro communication technology. Currently, we are working on the latter 
option. A report based on a full picture of DED is expected by the end of 2013.
\section{Related Work}
\label{Sec:RelWork}

The common model for event detection in wireless sensor networks (WSNs) is that each node relays
all of its locally generated data to the base station without local processing~\cite{GM:04}. 
This model works well for small-scale networks, a small amount of data per event, and lower rates of events per node. This model is inefficient for large-scale networks, as aggregated data
transmissions can easily exceed the available bandwidth en route to the central station.

Another model for event detection involves in-network processing. In this model, processing is done 
by the nodes to compute events of interest against certain criteria known to the nodes. 
This may significantly reduce the amount of communication and, hence, the energy consumed. 

In TAG~\cite{MFHJH:02} data is processed along the routing path at its intermediate nodes. This approach works
well for computing aggregates like $\textit{max}$, $\textit{min}$, $\textit{count}$, and $\textit{sum}$
etc. However, it is not suitable for event-detection schemes where acknowledging the node about detection of
event of interest is mandatory.

In \cite{MS:06}, the network is divided into equally sized cells.  
Each cell has a leader. Nodes within the cell route their data to a leader. 
The leader processes the data and informs other nodes in the cell about the 
decision. This scheme is scalable and also satisfies the requirement of 
acknowledging a node about the decision.
However, this scheme will easily produce false negatives in case an event 
occurs on the border of two or more cells.

In \cite{WDWS:10}, Wittenburg et al. present a distributed event detection 
scheme where each node decides for itself based on information from its 
neighbors. Their scheme scales well.  The node is also aware of the final 
decision made about the occurrence of an event. However, they do not explore 
the tradeoffs associated with this scheme under a high frequency of local events.

In \cite{WAJRLW:05}, Werner-Allen et al. deal with handling high rates of events up to 102 Hz per
node. They use a distributed event detection scheme that works as follows.  When a node triggers a local
event, it broadcasts a vote message. If any node receives enough votes from other nodes during some time
window, it initiates global data collection by flooding a message to all nodes in the network. This scheme
implies that when an event occurs it is detected by \emph{all} nodes in the network. However, based on this
assumption, the scheme is not scalable.

Finally, Werner-Allen et al. argue in \cite{WALJW:06}, to use a WSN as a scientific instrument.  They deal
with an event rate of 100Hz per node and high resolution data. Because of the high data rates it is
infeasible to transmit all sensor data.  So nodes locally detect interesting events and only transmit data
related to interesting events. They use an event detection algorithm that is basically centralized.  When a
node triggers an event, it is transmitted to the base station.  If the base station receives triggers from
$30\%$ of the active nodes within a 10 second window, then it is considered as an event of interest and data
collection is initiated. However, this scheme of event detection is not scalable for large geographical areas
and only applicable to specific domains, with network-wide events.

Concluding, the requirements that we mentioned have been partially addressed by a multitude of
schemes. However, to the best of our knowledge, no work lies in the intersection of these areas to propose a
scalable solution and address the design tradeoffs for the applications class we target at under the
constraints of limited energy and capacity of the WSN nodes.

\section{Conclusion and Future Work}
\label{Sec:Con}

We notice that our proposed distributed event detection algorithms use information only from geographical
neighbors and perform analysis locally at the station to discard irrelevant data.  In contrast, a centralized
approach requires each station to send its data (N1 triggers) to the CRS for analysis.  This implies that as
the number of hops between the station and the CRS increases, the communication cost of sending the N1
triggers to the CRS also increases. 
We see in Figure~\ref{bw_quan_small} that for a network of size $\netsize=91$, with average hop-distance $\overline{h}=3$ and maximum hop-distance $h_{max}=5$ from the CRS, the distributed approach outperforms the centralized.
This means that the centralized approach is not geographically scalable.
On the other hand, our algorithms enable a station to analyse data locally and show strong affinity for
geographical scalability.

In this paper we focused on exploring the possibility of collaborative local data analysis to build
geographically scalable solutions for ultra-high energy comic ray detection. The devices used for \emph{cosmic
  ray air shower} detection have limited communication, processing, and storage capacity. Each detector also
has a limited energy budget.

Wireless communication among cosmic-ray detectors is the only way to cover a large spatial area for detecting
cosmic-ray air showers.  Each cosmic-ray detector generates a huge amount of data that needs to be sent to the
CRS. But wireless communication offers limited bandwidth that does not meet the bandwidth requirements
of a centralized solution for this application.

We present a distributed event detection algorithm that enables a cosmic ray detector to analyse its data
locally based on information from its geographical neighbors. The ability of collaborative local data
analysis makes our algorithm attractive for building large geographically scalable solutions. The results from
simulating our algorithm show that a significant amount of irrelevant data can be filtered out locally. This
allows a detector to utilize wireless communication bandwidth more efficiently. We also explore several design
tradeoffs that help understand the relationship between accuracy of our algorithm and various resources usage.

We know that wireless communication medium has not only limited amount of bandwidth but wireless 
communication links are also unreliable. In this paper, we assume that links are reliable. As future work we will explore the behaviour of our distributed event detection algorithm under unreliable communication channels.
\section{Acknowledgements}
\label{Sec:ack}
We thank the Pierre Auger Collaboration for using part of their data sets and stimulating discussions with C. Timmermans and A.M. van den Berg in an early stage of the project.


\end{document}